# Highly Rectifying Cubic Copper Iron Sulfides p-n Junction Diode Fabricated by Anodic Oxidation

*Yoshimine Kato\*, Tomoaki Nakamura, Katsuya Komorita, and Kungen Teii*

Y. Kato, ORCiD: 0000-0001-6804-0285, T. Nakamura

Department of Automotive Science, Graduate School of Engineering

Kyushu University

Nishiku, Fukuoka 819-0395, Japan

Email: yoshimine.kiho.kato@kyudai.jp

Komorita, ORCiD: 0009-0008-7026-1739

Department of Materials Science and Engineering,

Kyushu University,

Nishiku, Fukuoka 819-0395, Japan

K. Teii, ORCiD: 0000-0003-4090-5376

Department of Advanced Energy Science and Engineering,

Kyushu University,

Kasuga, Fukuoka 816-8580, Japan

**Funding:** There is no funding for this research.

**Abstract**

Rectification properties of semiconductor p-n junction diodes are the basic and important characteristics for electronic device evaluation, especially for novel semiconductor materials. Today's semiconductor devices' fabrication and integration processes require multibillion-dollar investments and are desired to be reduced or simplified. Therefore, low-cost and non-toxic base metal materials with simple fabrication methods are desired for the future semiconductor industry. Recently, copper-based sulfides have been studied for semiconductor devices such as thermoelectric, photovoltaic, or water-splitting applications. Here, a highly rectifying p-n diode of a cubic (disordered) phase $Cu_4Fe_5S_8$ polycrystal with a zincblende-like structure fabricated by a simple/low-cost wet process is shown. It is found that the $Cu_4Fe_5S_8$ diode shows the highest rectification ratio in the order of $10^6$ with a large forward current density of 15 $Acm^{-2}$ (@1.5 V forward bias) at room temperature among the other compounds of copper iron sulfide devices. This remarkable and stable diode characteristic obtained by the p-type layer anodically grown on the sintered n-type cubic-$Cu_4Fe_5S_8$ can bring the industry closer to low-cost semiconductor manufacturing. These results open a platform of novel semiconductor materials such as cubic-$Cu_4Fe_5S_8$ with further superior crystal growth and conductive characteristics.

# 1. Introduction

The Cu-Fe-S system, as represented by chalcopyrite ($CuFeS_2$), is a well-known copper mineral. Chalcopyrite-containing ores are the world's most important natural resource for copper production, and their extraction process is an important technological innovation.[1-3] Besides the conventional pyrometallurgical process, a number of leaching methods on chalcopyrite and its-containing ores have been studied.[4-8] During the leaching process, surface layers that hinder the oxidative dissolution rate of the chalcopyrite mineral were formed. Ghahremaninezhad et.al.[5] analyzed the passive layers during the leaching process with the anodic dissolution of $CuFeS_2$ in $H_2SO_4$ (0.5 M) electrolyte using Mott-Schottky techniques. It has been shown that the passive film of chalcopyrite has semiconductive properties. At higher electrode potentials (around 0.75 V vs SHE (saturated hydrogen electrode)), the formation of CuS with $Fe_2(SO_4)_3$ concomitant was found. n-type or p-type conduction can be distinguished from the slope of $C^{-2}$ curves of Mott-Schottky plots,[9, 10] where $C$ is the capacitance density of the specimen.

Parker et. al. studied the oxidative leaching of copper from chalcopyrite, and referred to the intermediate product as a p-type CuS having semiconducting properties.[11] Warren et.al. reported the electrochemical oxidation of $CuFeS_2$ in $H_2SO_4$ (0.01 M, 0.1 M, and 1 M) electrolytes.[7] During the electrochemical reactions, the applied voltage and the formation of intermediate products are found to be important. At the higher potential range the current increases due to the release of the soluble $SO_4^-$, $Cu^{+2}$, and $Fe^{+3}$ by the decomposition of water forming chemisorbed oxygens. Nava et.al. also studied the anodic dissolution of chalcopyrite in $H_2SO_4$ (1.7 M) electrolytes with carbon paste electrodes.[12] Iron has been completely released at the higher potential region of 1.085 V - 1.165 V vs SHE, and the formation of covellite (CuS) was identified. Vaughan et.al. reported that the electrochemical oxidation in $HClO_4$ (1 M) at the electrode potential region of 0.2-0.6 V vs. SCE (saturated calomel reference electrode) at 298 K of chalcopyrite ($CuFeS_2$) and its metal-enriched derivatives like haycockite ($Cu_4Fe_5S_8$), the oxidation products are $Cu_xS_y*$, $Fe^{3+}$ ions at electrode potentials > 0.53 V vs. SCE on the electrode surface.[13] They also reported that the order of relative oxidation rates is haycockite > mooihoekite > talnakhite > chalcopyrite.

Many types of chalcopyrites and their related structure materials such as II-IV-$V_2$ group and I-III-$VI_2$ group compounds have also been studied for functional semiconductors. For example, Cu(In, Ga)$Se_2$ (CIGS) or $CuInSe_{2-x}S_x$ (CIS) for photovoltaic devices have been studied.[14] These materials have also been applied to photoelectrochemical water splitting.[15] On the other hand, device applications of $CuFeS_2$ have been made and reviewed for thermoelectric, photodetection, and photovoltaic materials.[16] A number of previous studies on Cu-Fe-S materials, such as chalcopyrite ($CuFeS_2$) show that they have semiconductor characteristics.[5,6,8,17,18] Tsujii et.al. showed that $Cu_{1-x}Fe_{1+x}S_2$ with x =0.03 and 0.05 have thermoelectric properties.[19] Xie et.al. studied $CuFeS_{2+2x}$ and talnakhite $Cu_{17.6}Fe_{17.6}S_{32}$ thermoelectric materials.[20,21] These Cu-Fe-S compositions are relatively close to $Cu_4Fe_5S_8$ in our study. There was a report on $Cu_4Fe_5S_8$ used as a catalyst for water splitting.[22] It was found that the optical band gap is 1.26 eV for haycockite ($Cu_4Fe_5S_8$).[23] A large number of studies have shown the n-type and p-type layer formation on the Cu-Fe-S materials, however, there were only a few reports on p-n junction devices like nanocrystal $CuFeS_2$/n-Si heterojunction photodetectors,[24] and no reports on chalcopyrite or $Cu_4Fe_5S_8$ pn-(homo-)junction diode devices. The present work demonstrates that the surface layer grown on the sintered n-type bulk $Cu_4Fe_5S_8$ substrates by the anodic oxidation wet process works as a p-type layer, and that the fabricated p-n junction device shows highly rectifying characteristics.

Table 1.　Properties of prepared pellet samples.

| Sample No. | Preparation composition | Density, $d$ [g cm$^{-3}$] | Relative density, $R$ [%] | Resistivity $\rho_v$ [Ω cm] | Conduction type |
|---|---|---|---|---|---|
| A | $Cu_{0.235}Fe_{0.294}S_{0.471}$ ($Cu_4Fe_5S_8$) | 3.85 | 96[a] | $3.5 \times 10^{-3}$ | n |
| B | $Cu_{0.25}Fe_{0.25}S_{0.50}$ ($CuFeS_2$) | 4.06 | 96[b] | $1.6 \times 10^{-2}$ | n |

[a] Calculated based on the theoretical density (4.00 g cm$^{-3}$) evaluated from its nominal composition and estimated crystal structure(cf. Figure 1).

[b] The relative density calculated based on the theoretical density of chalcopyrite (4.18 g cm$^{-3}$, PDF #00-037-047) was 96-97 %.

## 2. Results and Discussion

### 2.1. Crystallographic Characteristics

**Table 1** shows the analysis results for the density and conduction of the pellet samples (substrate material) synthesised by the pulsed current pressure sintering (PCPS) method followed by annealing. The composition of raw materials of sample A is the stoichiometric one of haycockites ($Cu_4Fe_5S_8$ or $Cu_{0.235}Fe_{0.294}S_{0.471}$), and that of sample B is the stoichiometric one of chalcopyrites ($CuFeS_2$ or $Cu_{0.25}Fe_{0.25}S_{0.50}$). The relative density of both samples is high enough so that a quite flat surface was observed after the polishing. The resistivity and conduction types were measured using the four-point probes and the Seebeck effect measurements, respectively. It was found that both substrate samples are n-type, and the resistivity of sample A is about 3.5 m Ωcm, which is about one order of magnitude lower than that of sample B.

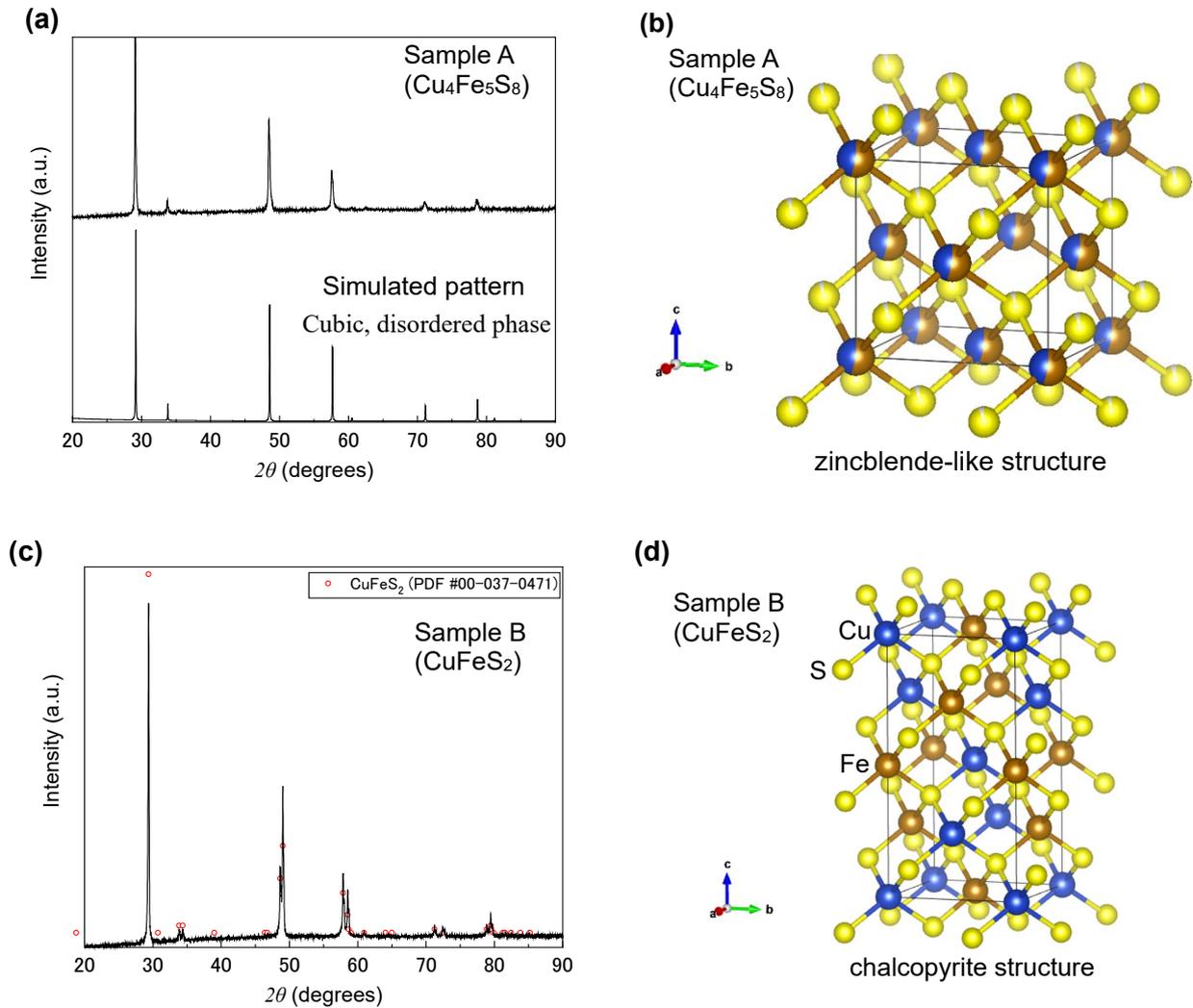

**Figure 1.** X-ray diffraction (XRD) (Cu-Kα) patterns and crystal structures of synthesized pellets. a) Sample A with a nominal composition of $Cu_4Fe_5S_8$ shows a zincblende-like cubic disordered phase. A simulated XRD pattern from the cubic-$Cu_4Fe_5S_8$ crystal structure shown in (b) is at the lower part of the figure, which matches well with each other. b) Crystal structure of sample A showing cubic-$Cu_4Fe_5S_8$ disordered structure F-43m with a lattice constant of 0.53 nm. Sites of Cu and Fe cannot be determined accurately by XRD. c) Comparison of experimental XRD data of sample B that shows a typical chalcopyrite single phase. d) Crystal structure of sample B showing typical I-42d $CuFeS_2$ tetragonal structure. Blue spheres are Cu, brown spheres are Fe, and yellow spheres are S.

Crystallographic structures of substrate samples A and B were investigated with X-ray diffraction (XRD) as shown in **Figure 1a,c**. The XRD pattern of sample A does not match that of the haycockite ($Cu_4Fe_5S_8$: PDF #04-009-1412; orthorhombic).[25-27] Sample A has a cubic disordered zincblende-like structure with a lattice constant of 0.53 nm and nominal composition of $Cu_4Fe_5S_8$ as shown in the simulated XRD pattern at the bottom of Figure 1a and the schematic drawing of the crystal structure F-43m in Figure 1b. On the other hand, Sample B in Figure 1c shows a chalcopyrite structure (PDF #00-037-0471; tetragonal)[28] with a nominal composition of $CuFeS_2$. A typical chalcopyrite I-42d $CuFeS_2$ tetragonal structure is shown in Figure 1d. In this study, the pellet sample was annealed at 500 °C in order to homogenize its microstructures after sintering. It is considered that the high-temperature disordered structure is stable at this annealing temperature, and the atomic arrangement is sufficiently regularized during the cooling process. Since the majority of the III-V semiconductors such as GaAs, InP, and GaP have zincblende-like structures, this crystal structure could also be the reason for the remarkable semiconductor characteristics of sample A as shown later. Since this lattice constant is close to the Si lattice constant of 0.543 nm, it may be possible to grow $Cu_4Fe_5S_8$ film epitaxially on a Si (100) substrate. [29-31]

## 2.2. Mott-Schottky plots

The sintered Cu-Fe-S substrates were anodically oxidized in sulfuric acid (0.18 M) solution to form p-n junctions on their surfaces. Anodic oxidation was performed by the Mott-Schottky technique. That is, the current and capacitance were measured while sweeping the electrode potential. During the anodization, Mott-Schottky plots were used to confirm that the carrier type at the topmost surface was p-type. **Figure 2a,b** show the Mott–Schottky plots for samples A and B at frequencies of 100 Hz and 1 kHz, respectively. In the case of chalcopyrite, it is reported that a few types of dissolution reactions proceed during the anodic oxidation.[7,11-13] This reaction depends on the potential and the formation of an intermediate layer on the electrode surface. For both samples A and B, as the potential increases, the curves of $1/C^2$ show peaks at 0.9 to 1.1 V in the Mott-Schottky plot. The slope of $1/C^2$ curves is positive at a lower potential, showing n-type conduction. On the other hand, at some peak voltage $Vp$, the slope becomes negative due to opposite charge carrier conduction. This suggests the formation of a p-type reaction intermediate layer at the solution/electrode interface.[5]

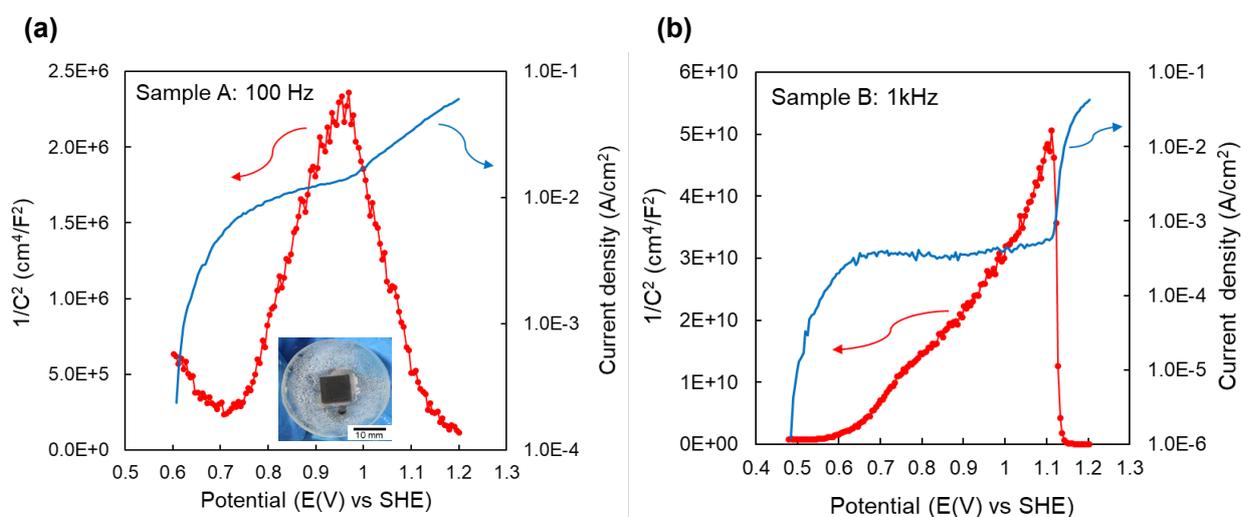

**Figure 2.** Mott-Schottky plots and current-voltage curve in $H_2SO_4$ (0.18 M) at 70 °C. a) Sample A measured at 100 Hz showing the p-type layer growth after the peak of the $1/C^2$ curve around 0.95V vs SHE. A picture of the sample embedded in resin is inserted. b) Sample B measured at 1 kHz shows quick growth with less carrier density of the p-layer after the peak.

Crundwell et.al. have reported for the $CuFeS_2$ that non-stoichiometric dissolution occurs with more rapid removal of iron than copper from the lattice during the formation of the p-type layer in the solution.[6] Also, above $Vp$, the oxidation current increases rapidly and then saturates afterwards. It has been reported that the same phenomenon occurs in the dissolution of $CuFeS_2$ in $H_2SO_4$ (0.3 M) electrolyte at 20 °C using the Mott-Schottky plot and explained the mechanism with band diagrams.[32] Warren et.al. have reported that during the electrochemical reactions of $CuFeS_2$ at the higher potential range, the current increases due to the release of the soluble $SO_4^-$, $Cu^{+2}$, and $Fe^{+3}$ by the decomposition of water-forming chemisorbed oxygens.[7] Previous researchers have reported that during the electrochemical anodic oxidation of chalcopyrite or haycockite ($Cu_4Fe_5S_8$) at higher electrode potential, the intermediate layer could be the semiconducting p-type CuS layer.[5-7,13] CuS has been reported as a p-type semiconductor with copper ions as carriers in the lattice.[33,34] The formation of the p-n junction by the anodization was confirmed by the following rectification measurement experiment.

The carrier density of the sintered pellets was calculated from the slopes of Mott–Schottky plots in Figure 2a,b. Since the Mott-Schottky model is based on the band structure of a metal-semiconductor junction,[35,36] the carrier concentration of the metal is considered ∞ and can be neglected according to the capacitance calculation.[37] Therefore, the carrier concentration of sample A's pellet substrate is normally calculated to be too high in the order of $10^{21}$ cm$^{-3}$ (degenerate type), however, this value can also be possible according to the prediction of Miller et al.[38] Since the carrier concentration in the H$_2$SO$_4$ (0.18 M) electrolyte is calculated as the SO$_4^-$ (or H$^+$) concentration of $1.08 \times 10^{20}$ cm$^{-3}$,[39] we consider that these carrier concentrations of SO$_4^-$ (or H$^+$) in the electrolyte are not as high as the electrons in the metal. The details of this calculation method using the capacitance of the semiconductor p-n anisotype heterojunction are described in the Experimental Section. In this study, the donor concentration of the sintered pellet substrate $N_D$ was calculated from the positive slope value to be about $4.3 \times 10^{19}$ cm$^{-3}$. Since the negative slope for the p-type layer of sample A in Figure 2a is about the same as the positive slope, the acceptor concentration $N_A$ is about the same as $4.3 \times 10^{19}$ cm$^{-3}$. On the other hand, the negative slope of sample B is steeper than that of the ones shown in Figure 2b. According to Equation (1) in the Experimental Section, the carrier concentrations were calculated as $N_D = 6.8 \times 10^{17}$ cm$^{-3}$ for the n-type substrate and $1.2 \times 10^{18}$ cm$^{-3}$ for the p-type layer grown on the surface. This lower carrier concentration may be related to the inferior current-voltage (*I-V*) characteristics of sample B (Figure S1a, Supporting Information) and thinner p-type layer thickness due to the lowest oxidation rate of chalcopyrite as suggested by Vaughan et.al.[13]

## 2.3. Morphological Characteristics

SEM images of the surface morphology of sample A before and after the anodization are shown in **Figure 3a,b**. After polishing the surface of the samples with #2000 SiC polishing paper, the morphology of sample A shows a flat surface with small voids (Figure 3a). After the anodization, the surface becomes significantly rough, as seen in Figure 3b. An increase in the number of larger voids suggests that some particles may partially dissolve. For comparison, SEM images of the surface morphology of sample B are shown in Figures S1c, d (Supporting Information). Even after the anodization, the smoothness of the surface is almost retained for sample B. As reported by Vaughan et.al.[13], this may be because sample B, chalcopyrite, shows a lower degree of oxidative dissolution compared to sample A, as shown by the lower anodic current density in the potential range lower than $V_p$ in Figure 2b. Moreover, the X-ray spectroscopy (EDS) mapping after the anodization confirmed that the in-plane uniformity of the stoichiometric composition was very high.

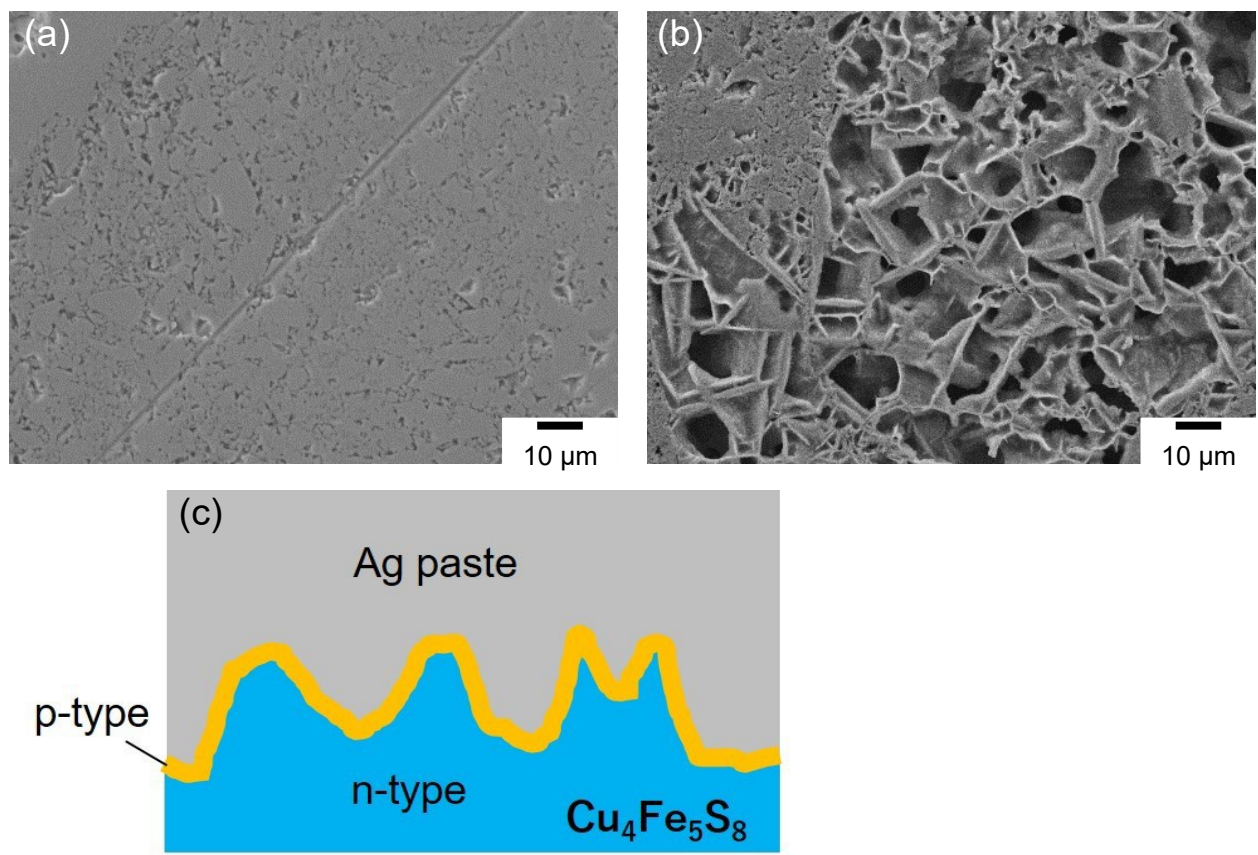

**Figure 3.** SEM images of the surface morphology of sample A (cubic-$Cu_4Fe_5S_8$). a) Before the anodization, many small pits appear after the polishing. b) After the anodization, the p-n junction is formed even on this rough surface. c) Schematic drawing of the cross-section of the p-n junction showing a remarkable growth of p-type layer on the rough surface after the anodization. Ag silver paste is coated on the surface as an electrode. A different magnification of the surface morphologies of sample A is shown in Figures S1a,b (Supporting Information).

In Figure 3c, a schematic drawing of the p-type layer, which is naturally formed on the n-type pellet substrate after anodization, is shown with an Ag paste electrode on the top. It is important to mention that even with the rough surface of sample A, the p-n junction was likely formed as depicted in Figure 3c because excellent *I-V* characteristics were obtained as shown later. Such good coverage of the p-type layer on the rough n-type substrate using the anodization technique can be quite useful for large-scale integrated (LSI) semiconductor processes such as trench structures.

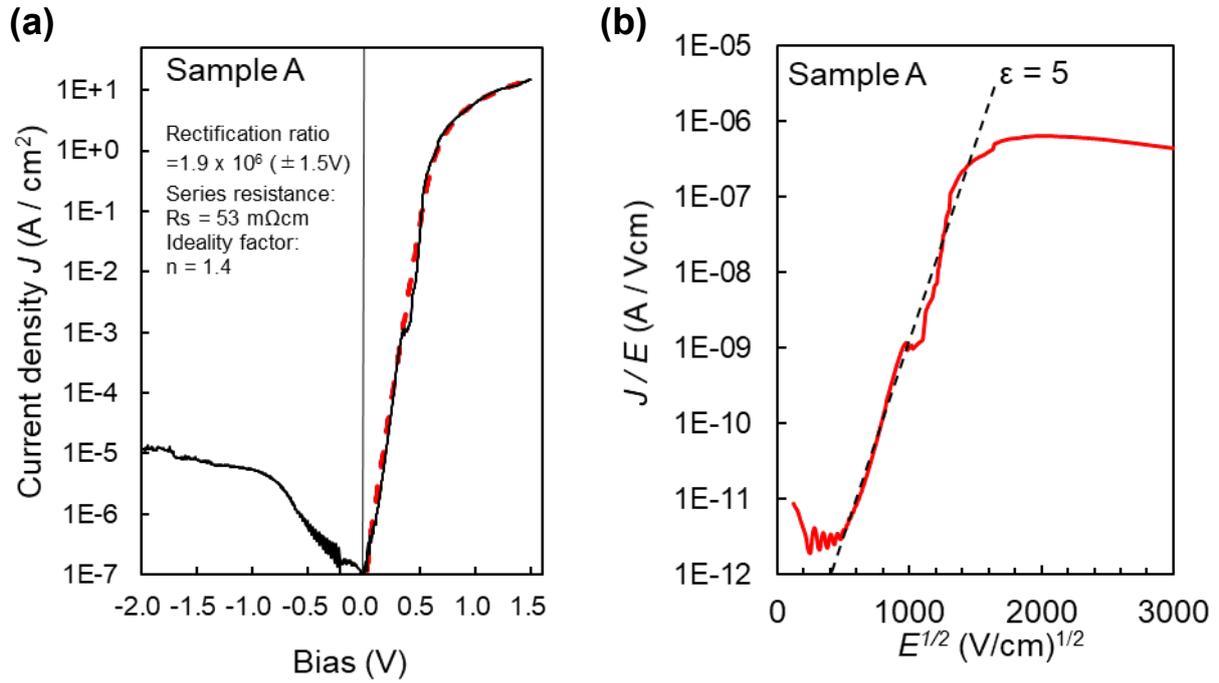

**Figure 4.** Current-voltage (*I-V*) characteristics of the p-n junction of sample A measured at room temperature (RT). a) Log plots of current vs applied bias showing a rectification ratio of about $1.9 \times 10^6$ (at ±1.5V) and forward current achieving nearly 15 A/cm² at 1.5 V. The red dashed line shows a fitting curve according to Equation (2) in the Experimental Section. A series resistance and an ideality factor were obtained from the fitting. A slope of forward bias shows the ON resistance of $10^{-8}$ Ω order. b) The plot of the forward current density $J$ divided by applied field $E$ vs the square root of the applied field ($E^{1/2}$) becomes nearly straight in the forward bias region, indicating the current transport is limited by the Frenkel-Poole conduction. The slope of the plot matches that of a dashed line corresponding to the theoretical slope for the relative permittivity $\varepsilon_s = 5$. The built-in potential $V_b = 0.9$ V was used for this calculation.

## 2.4. Electrical Characteristics

**Figure 4a** shows the *I-V* characteristics of the sample A p-n junction diodes measured at room temperature. Since the reverse current densities are quite low as shown in Figure 4a, the highest rectification ratio reaches 1.9 x10$^6$ (at ±1.5V) with the forward current of about 15 A/cm$^2$ at 1.5 V, which is as high as conventional solid-state devices.[37,40,41] *I-V* characteristics of sample B are shown in Figure S2 (Supporting Information). A large forward current density is observed even at the forward bias voltage of 0.5 V, however, the reverse current density is quite high; therefore, the rectification ratio was only about 50 (at ±1.0V). This large reverse leak current was also observed in other Cu-Fe-S system samples.

A red dashed line shows the typical curve fitting of the *I-V* characteristics in the forward bias region by using the theoretical p-n junction *I-V* Equation (2) in the Experimental Section. The series resistance: Rs of 53 mΩcm and the ideality factor: n of 1.4 were obtained by the fitting with the red dashed line as shown in Figure 4a. This series resistance is consistent with that measured by the four-point probes method for the substrate of sample A and is quite low compared with the common Si p-n devices.[42] Therefore, the forward current at 1.5 V is quite high, and a slope of forward bias shows the ON resistance of 10$^{-8}$ Ω order. The noise error of the current density in the lower voltage region is due to the noise from the measurement system. In addition, little or no photovoltaic characteristic was observed under AM1 illumination.

Figure 4b shows the plots of *J/E* versus *E*$^{1/2}$ for sample A measured at RT. The plot of the forward current by assuming the Frenkel-Poole conduction[37] that is typical for disordered materials shows a nearly straight line as indicated by a dashed line in Figure 4b. From its slope, the relative permittivity of sample A, *ε$_s$* ≃ 5 is derived from Equation (3) in the Experimental Section. This value is almost consistent with the value of chalcopyrite at a frequency of 1 GHz and the optically measured one. [32,43]

It is worth mentioning that, even using 3N or 4N purity starting materials with this wet process for fabricating a p-n junction, such high rectification with large forward current density is obtained. Additionally, the *I-V* and *C-V* data were repeatable even after several months of storage in a desiccator, indicating the high durability and stability of the devices. Therefore, it would be possible to improve and stabilise the $Cu_4Fe_5S_8$ p-n diode's performance using higher-purity materials and electrolytes. In comparison with Si device technology, Si starting material needs to have very high purity such as 11N to have stable device performance. In addition, Si is readily oxidized, and therefore it is necessary to avoid contact with air as much as possible, making the process difficult and costly. Since there is no study on the natural oxidation of cubic-$Cu_4Fe_5S_8$ yet, we estimated from the study of the natural oxidation reaction of $FeCuS_2$ at RT in air.[44] The standard Gibbs free energy of formation ($\Delta G_f°$) can be calculated to be much higher than that of Si oxidation. Therefore, it is expected that $Cu_4Fe_5S_8$ is oxidized more slowly compared with Si, and high reliability and long life are also expected when it is applied to devices. In addition, Fe, Cu, and S elements are rich in resources and inexpensive. Thus, the device processing and packaging can be expected to be much easier without large concern for high purity and cleanliness than Si devices, and manufactured at a low cost.

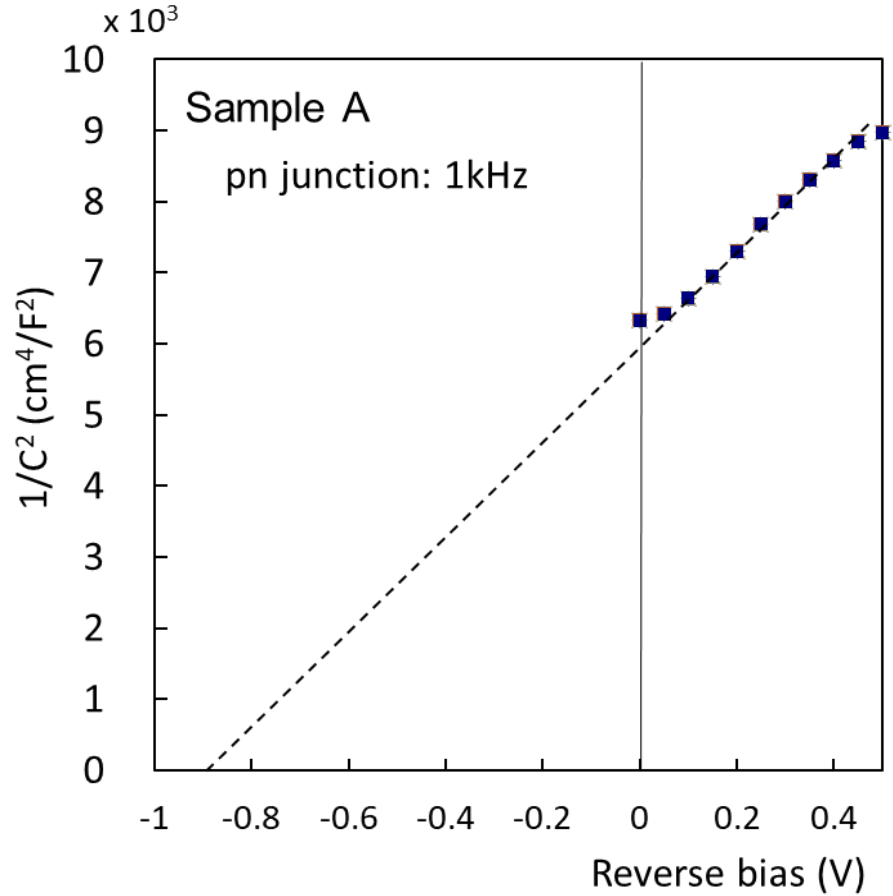

**Figure 5.** $1/C^2$ - $V$ plot of the p-n junction of sample A measured at 1 kHz. An extrapolation line (dashed line) to estimate the diffusion potential of the p-n junction is about 0.9 eV. The carrier concentration calculated from the slope matches the one calculated from the Mott–Schottky plots in Figure 2a.

A plot of $1/C^2$ vs applied bias voltage $V$ of the sample A p-n junction in **Figure 5** shows the builtin potential $Vb$ of about 0.9 eV according to Equation (1) in the Experimental Section. Since the band gap of haycockite is estimated to be about 1.26 eV,[23] this $Vb$ value is reasonable and large enough to cause high rectification. As the permittivity of $Cu_4Fe_5S_8$ with zincblende structure is not available in the literature, the permittivity of 70 at 1 kHz for chalcopyrite structure was used for sample A for calculating $N_D$.[43,45] In Equation (1) of the Experimental Section, the relative permittivity of the semiconductor, $\varepsilon_E$, and the $H^+$ concentration in the electrolyte $N_E$ can be simply replaced as the p-type layer's $\varepsilon_s$ and $N_A$, respectively. Since n-type and p-type are the same material, it is considered as $\varepsilon_s = \varepsilon_E$. Then, the doping concentration in p-type layer $N_A$ is calculated from the slope of Figure 5 using Equation (1) as $N_A = 4.3 \times 10^{19}$ cm$^{-3}$. This value is the same as the substrate bulk carrier concentration, $N_D = 4.3 \times 10^{19}$ cm$^{-3}$ calculated from the Mott–Schottky plot as previously derived. This suggests that the n-layer and p-layer are originally the same material which is not doped intentionally with other impurities. However, the mechanism of p-type conduction of the intermediate layer after the anodic dissolution process needs more study. The surface area can also be calculated from the plot of $1/C^2$ vs $V$ according to Equation (1). Due to the rough surface of sample A, the surface area is calculated to be about 500 times larger than that of the electrode size. As shown in Figure 2c, by using this anodization method, the p-n junction with micron-order irregularities on the surface can be spontaneously formed. This fact also leads to an increase in the surface area of more than several hundred times, thus dramatically improving the performance of semiconductor devices.

## 3. Conclusion

We have demonstrated the semiconducting characteristics of the cubic-$Cu_4Fe_5S_8$ polycrystal applicable for highly rectifying p-n junction diodes. The p-type layer was formed on the synthesized n-type cubic-$Cu_4Fe_5S_8$ bulk substrate via anodic polarization in the 0.18 M sulfuric acid solution for the first time. The highest $I$-$V$ rectification ratio achieved was in the order of $10^6$ with a large forward current of 15 A/cm$^2$ at 1.5 V forward bias. The XRD analysis showed that the synthesized $Cu_4Fe_5S_8$ polycrystal substrate had a disordered cubic zincblende-like structure with a lattice constant of 0.53 nm, unlike a typical haycockite (orthorhombic) structure. According to the $I$-$V$ and $C$-$V$ characteristics, the carrier concentrations of the n-type substrate and p-type layer of the cubic-$Cu_4Fe_5S_8$ were almost the same and calculated to be about $4.3 \times 10^{19}$ cm$^{-3}$. In addition, $\varepsilon_s \simeq 5$ (@ high frequency) was calculated. As far as we know, there were no reports on the $Cu_4Fe_5S_8$ p-n junction diode device made by an anodic wet process, which enables revolutionary cost reductions in future semiconductor manufacturing processes. In addition, Fe, Cu, and S elements are rich in resources and inexpensive. Therefore, this leads to a large cost-down, high reliability, and long-life devices in semiconductor fabrication processes. Further studies on electronic band structure, defects, and transportation mechanisms are necessary for this cubic-$Cu_4Fe_5S_8$ and other Cu-Fe-S systems[46] which may have higher potential and advantages as electronic devices.

## 4. Experimental Section

*Synthesizing and analyzing crystals:* The bulk $Cu_4Fe_5S_8$ was first synthesized in a vacuum-sealed quartz ampule by the direct reaction of Cu (99.96%), Fe(99.5%) wires, and S (4N) powder. Each material's weights were adjusted to match the $Cu_4Fe_5S_8$ stoichiometry. The ampule was heated at 520 °C for 48 h. Secondly, the synthesized bulk was ground in a mortar under an air atmosphere, pressed into pellets, and then heated in an evacuated quartz ampule in a vacuum at 520 °C for 24 h. The pellet was ground again in the air and sintered by the PCPS method at 40 MPa in Ar at 560 °C for 10 min. After sintering, the pellet sample was annealed in a vacuum at 500 °C to homogenize its microstructure. There was no intentional doping during the process. The samples were characterized by powder XRD to examine the crystal structures of each sample. Surface morphology and chemical compositions were analysed using an electron microscope equipped with energy-dispersive EDS analysis. XRD and EDS analysis showed the crystal structures and chemical compositions of the fabricated samples as either cubic-$Cu_4Fe_5S_8$, chalcopyrite ($CuFeS_2$) (CP), mooihoekite ($Cu_9Fe_9S_{16}$) (MH), talnakhite ($Cu_9Fe_8S_{16}$) (TK), cubanite ($CuFe_2S_3$) (CN), and bornite ($Cu_5FeS_4$) (BN). The electrical characteristics of the fabricated HC bulk crystal were measured using the four-point method and the Seebeck effect measurement using needle-shaped electrodes.

*Mott–Schottky method*: Electrochemical measurements were carried out with a conventional three-electrode electrolytic cell using a platinum wire as a counter electrode. The working electrode (WE) assembly was done by mounting the cut-out 7 x 7 x 3 mm Cu-Fe-S pellet substrate in non-conductive epoxy resin as a structural support. Prior to the test, the substrate was polished with #2000 SiC polishing paper. The Counter and reference electrodes were Pt wire and Ag-AgCl electrode, respectively. [5,6,47,48] A Pt film was sputtered on the back side of the substrate as a back Ohmic contact, and a copper wire was attached using the Ag paste (DOTITE, AA55, Fujikura Kasei). A sulphuric acid (0.18 M) solution was used as the electrolyte at 70 °C and stirred at 400 rpm. To form a p-type layer on the polished bulk substrate (i.e., WE), it was anodically polarized from the resting potential using the Mott-Schottky technique. To measure the AC impedance and current a potentiostat (Versa STAT 3F, Princeton Applied Research) was used at the frequencies of 100 Hz, 1 kHz, and 10 kHz with an amplitude of 10 mV and a step of 5 mV from its rest potential with a sweep rate of about 1.2 s / step. The p-n junctions of cubic-$Cu_4Fe_5S_8$, CP, MH, TK, CN, and BN samples were also fabricated and measured in the same way. The carrier density of the sintered pellets was calculated from the slopes of the Mott–Schottky plots. The Schottky–Mott equation is based on the band structure of a metal-semiconductor junction. [35,36] However, when the electrolyte carrier concentration is not as high as the metal, we assume the capacitance of the semiconductor-electrolyte interface is similar to the capacitance of the semiconductor p-n anisotype heterojunction expressed in the following Equation (1) derived from Poisson's equation, [3,8,9]

$$\frac{1}{C^2} = \frac{2(\varepsilon_0 \varepsilon_E N_E + \varepsilon_0 \varepsilon_s N_D)}{q \varepsilon_0^2 \varepsilon_E \varepsilon_s N_E N_D} (V_b - V) \tag{1}$$

where $C$ is the measured capacitance per cm$^2$, $V$ is the applied electrode potential, $V_b$ is the built-in potential, $q$ is the elementary electric charge, $\varepsilon_0$ is the permittivity of free space, $\varepsilon_s$ is the relative permittivity of the semiconductor, $\varepsilon_E$ is the relative permittivity of the electrolyte, $N_D$ is the donner concentration of the sintered pellet substrate, and $N_E$ is assumed as the H$^+$ concentration in the electrolyte. Since the electrolyte carrier concentration of SO$_4^-$ (or H$^+$) H$^+$ concentration of H$_2$SO$_4$ (0.18 M) is 1.08 x 10$^{20}$ cm$^{-3}$, we assume that the carrier concentration in the electrolyte is not as high as the electrons in the metal. Since the concentration of H$_2$SO$_4$ is low, the relative permittivity of the electrolyte is assumed to be the same as water, 64 at 70 °C.[49,50] Since the permittivity of sample A, Cu$_4$Fe$_5$S$_8$, is not available in the literature, a value of chalcopyrite as 160 at 100 Hz was used for calculating $N_D$.[43,44] The $N_D$ was thus calculated from the positive slope value to be about 4.3 x10$^{19}$ cm$^{-3}$.

*p-n diode device characterization*: After the p-type layer was formed in the electrolyte on the n-type substrate, Ag paste electrodes of about 1 mm diameter were formed on the surface. *I-V* and *C-V* characteristics were measured at RT using a voltage and current source (Agilent B1505A). Capacitance was measured at 1 kHz, 10 kHz, 100 kHz, and 1 MHz. p-n junctions of cubic-Cu$_4$Fe$_5$S$_8$, CP, MH, TK, CN, and BN samples were also fabricated and measured in the same way.

The *I-V* characteristics in the forward bias region are fitted with the theoretical p-n junction *I-V* Equation (2):[40]

$$J = J_0\{(e^{\left(\frac{q(V-JRs)}{nkT}\right)}-1\} \qquad (2)$$

where $J_0$ is the reverse saturation current, $R_S$ is the series resistance, $q$ is the elementary electric charge, $k$ is the Boltzmann constant, and $n$ is the ideality factor.

The *I-V* characteristics were used to examine if the carrier transport process follows the

Poole–Frenkel conduction[37,51-53]

$$J \propto E\exp\left\{\frac{-q\left(\phi_B-\sqrt{\frac{qE}{\pi\varepsilon_0\varepsilon_s}}\right)}{kT}\right\} \qquad (3)$$

where $J$ is the current density, $E$ is the applied electric field, $\Phi_B$ is the barrier height, $q$ is the elementary electric charge, $\varepsilon_s$ is the high-frequency relative dielectric constant of the semiconductor material, $k$ is the Boltzmann constant, and $T$ is the temperature. The value of $E$ was calculated assuming the voltage was applied to the depletion layer of the p-n junction which varies with the voltage.

## Acknowledgements

We thank Y. Taninouchi, H. Nakano, and S. Oohue for their help and support in doing experiments.

L. Majula is also acknowledged for helping with measurements.

## Data Availability Statement

Source data supporting the findings of this study are provided with this paper and the Supporting Information.

**Supporting Information**

Supporting Information is available from the Wiley Online Library or the author.

**Conflict of Interests**

The authors declare no conflict of interest.

**Table of contents:**

**Highly Rectifying Cubic Copper Iron Sulfides p-n Junction Diode Fabricated by Anodic Oxidation**

A highly rectifying p-n diode of a cubic (disordered) phase $Cu_4Fe_5S_8$ polycrystal with a zincblende-like structure is fabricated. The $Cu_4Fe_5S_8$ diode shows the highest rectification ratio in the order of $10^6$ with a large forward current density of 15 A/cm$^2$ (@1.5V forward bias) at room temperature. The p-type layer is anodically grown in sulfuric acid solution on a sintered n-type cubic-$Cu_4Fe_5S_8$.

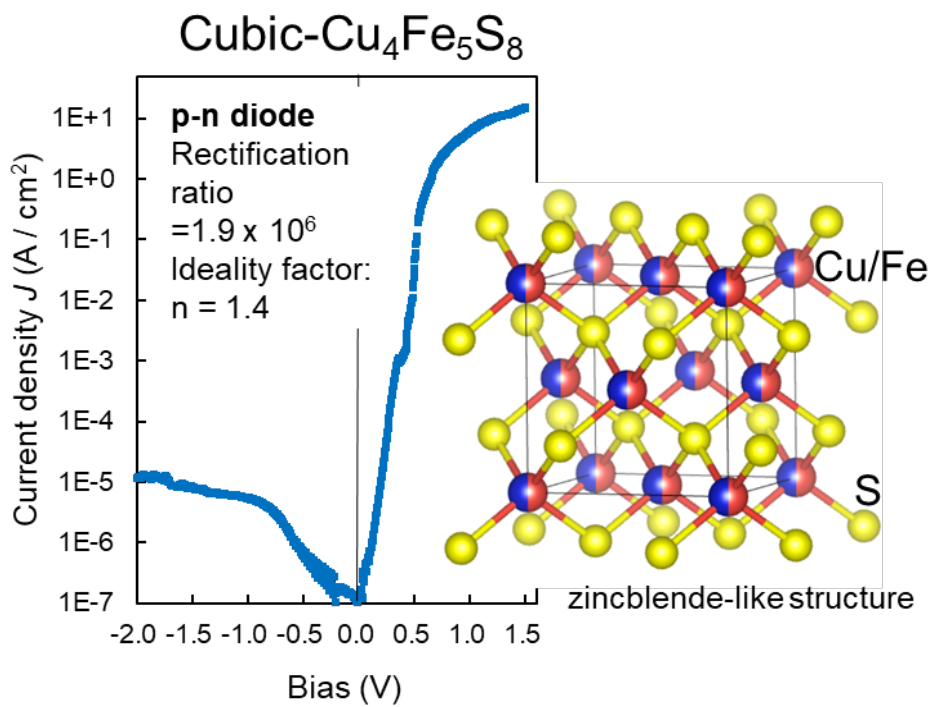